\documentclass[sn-nature]{sn-jnl}
\usepackage{times}
\usepackage{xcolor}
\usepackage[version=4]{mhchem}
\usepackage{amsmath,stmaryrd,graphicx,amssymb}
\usepackage{mathtools}  
\usepackage{xfrac}  
\usepackage{nameref}
\usepackage{miller}
\usepackage{xr}
\usepackage{gensymb}
\usepackage{nicefrac}
\emergencystretch=\maxdimen
\makeatletter
\renewcommand{\@seccntformat}[1]{%
  \ifcsname prefix@#1\endcsname
    \csname prefix@#1\endcsname
  \else
    \csname the#1\endcsname\quad
  \fi}
\newcommand\prefix@section{}
\newcommand{\prefix@subsection}{\thesubsection\ - }
\newcommand{\prefix@subsubsection}{\thesubsubsection\ - }
\renewcommand{\thesubsection}{\arabic{subsection}}
\makeatother
\newcommand{\quotes}[1]{``#1''}

\title[Article Title]{Direct determination of antiferroelectric-to-ferroelectric phase transition pathways in \ce{PbZrO3} with \textit{Operando} Electron Microscopy}

\author[1]{\fnm{Menglin} \sur{Zhu}}
\equalcont{These authors contributed equally to this work.}
\author[1]{\fnm{Michael} \sur{Xu}}
\equalcont{These authors contributed equally to this work.}
\author[2]{Louis Alaerts}
\author[3]{\fnm{Hao} \sur{Pan}}
\author[1]{\fnm{Colin} \sur{Gilgenbach}}
\author[2,4,5]{Geoffroy Hautier}
\author[4,5,6]{\fnm{Lane W.} \sur{Martin}}
\author*[1]{\fnm{James M.} \sur{LeBeau}}\email{lebeau@mit.edu}
\affil*[1]{\orgdiv{Department of Materials Science and Engineering}, \orgname{Massachusetts Institute of Technology}, \orgaddress{\city{Cambridge}, \postcode{02139}, \state{MA}, \country{USA}}}
\affil[2]{\orgdiv{Thayer School of Engineering}, \orgname{Dartmouth College}, \orgaddress{\city{Hanover}, \postcode{03755}, \state{NH}, \country{USA}}}
\affil[3]{\orgdiv{Department of Materials Science and Engineering}, \orgname{University of California, Berkeley}, \orgaddress{\city{Berkeley}, \postcode{94720}, \state{CA}, \country{USA}}}
\affil[4]{\orgdiv{Department of Materials Science and NanoEngineering}, \orgname{Rice University}, \orgaddress{\city{Houston}, \postcode{77005}, \state{TX}, \country{USA}}}
\affil[5]{\orgdiv{Rice Institute of Advanced Materials}, \orgname{Rice University}, \orgaddress{\city{Houston}, \postcode{77005}, \state{TX}, \country{USA}}}
\affil[6]{\orgdiv{Departments of Chemistry, and Physics and Astronomy}, \orgname{Rice University}, \orgaddress{\city{Houston}, \postcode{77005}, \state{TX}, \country{USA}}}
\date{}

\begin{document}

\maketitle 
\newpage
\begin{abstract}

Under a sufficiently high applied electric field, a non-polar antiferroelectric material, such as \ce{PbZrO3}, can undergo a rapid transformation to a polar ferroelectric phase. While this behavior is promising for energy storage and electromechanical applications, a complete understanding of the atomic-scale mechanisms governing the phase transition remain elusive. Here, we employ \textit{operando} scanning transmission electron microscopy electric field biasing to directly resolve the antiferroelectric-to-ferroelectric transition pathway in \ce{PbZrO3} thin films under device-relevant conditions. Atomic-resolution imaging reveals a multi-step transition that includes several metastable phases. Complementary nano-beam electron diffraction and atomic scale analysis further show that this pathway and its end states can be modulated, leading to the formation of a \quotes{dead layer} near the substrate with suppressed switching behavior. Taking advantage of this depth-dependent heterogeneity, dynamic phase transformations are observed between coexisting antiferroelectric and metastable ferroelectric phases. At this dynamic transition front, repeated phase interconversion is shown to be driven by competing internal (due to substrate clamping and extended defects) and external fields, allowing the relative energies of intermediate phases to be compared as a function of electric field. This work highlights the critical role of local energetics in phase stability and provides key experimental insights into field-induced phase transitions, guiding the design of antiferroelectric-based devices.

\end{abstract}

\newpage

\section{Main Text}

Antiferroelectric (AFE) materials are characterized by anti-parallel ordering of electric dipole moments, resulting in a non-polar global structure in the ground state. When subjected to a sufficiently high electric field, these materials can undergo a phase transition to a ferroelectric (FE) state with parallel aligned dipoles accompanied by significant changes in their functional properties \cite{Kittel1951-xp, Randall2021-xo, Zhuo2021-ad, Si2024-ix}. As a prototypical AFE material, lead zirconate (\ce{PbZrO3} PZO) exemplifies these unique characteristics \cite{Hao2014-ww}. For example, the rapid (de)polarization during the AFE-to-FE transition facilitates high-energy density and fast energy conversion \cite{Avdeev2006-mx, Dong2015-nt}. Furthermore, the AFE-to-FE transition is characterized by a large change in the primitive cell size that exhibits both temperature and strain sensitivity \cite{Liu2023-du, Pan2024-fr}, which is relevant for thermal regulation and electromechanical applications. These exceptional properties position PZO as a promising candidate for a diverse range of next-generation devices.

Beyond their technological significance, the phase transitions in PZO offer a platform to probe rich fundamental physics. The electric-field-driven AFE-to-FE transition, for instance, is theorized to emerge from competition between polar and antipolar phonon instabilities \cite{Xu2019-gw, Fu2022-vn, Zhang2024-nm, Waghmare1997-dc, Xu2024-aw}, potentially leading to transient bridging phases during switching \cite{Jiang2024-db, Wei2020-gu, Fu2022-vn}. These intermediate states, in turn, are thought to underpin the remanent ferroelectric ordering responsible for a slanted hysteresis loop deviating from ideal AFE behavior \cite{Fu2022-vn, Xu2019-gw, Si2025-cn}.

Detecting such transient states remains challenging in experiment due to their metastability, inhomogeneity in spatial distribution, and the timescales associated with their transitions \cite{Cai2003-mp, Blue1996-po, Wei2020-gu}. While strain \cite{Jiang2024-db} and chemical doping \cite{Fu2022-vn} have been used to stabilize intermediate phases for atomic-resolution (scanning) transmission electron microscopy (S/TEM), these stimuli have primarily functioned as proxies for probing electric-field biasing relevant to devices. In addition, while polarization switching is possible using electron-beam irradiation  \cite{Hart2016-ry,Wei2020-gu,Wei2021-mp, Calderon2023-cv}, the equivalent switching field becomes difficult to quantify, with irradiation artifacts such as ionization damage \cite{Egerton2004-mn} and local heating \cite{Kryshtal2022-pz} as well as irreversible transitions precluding direct comparison with device-scale response. Consequently, whether these phases are intrinsic to field-driven AFE–to-FE switching remains an open question.
% \cite{Cazaux1995-kq,Egerton2004-mn} also relevant for phase transitions?

Here, we employ \textit{operando} STEM to electrically bias PZO thin-film capacitors under device-relevant switching conditions. By implementing controlled field cycling and demonstrating electron-dose independence, we bypass beam-induced artifacts to isolate intrinsic structural dynamics in the thin films. Under applied electric fields, we show that the coupled order parameters defining the AFE ground state, antiparallel lead displacements and antiferrodistortive oxygen octahedral rotations, are simultaneously suppressed and compete with polar (FE) distortions. This competition subsequently gives rise to monoclinic intermediate phases, before favoring a rhombohedral and eventually tetragonal FE phase at high field. Finally, by leveraging epitaxial engineering, we show that internal fields from substrate clamping and dislocations, together with the external applied electric field, stabilize a dynamic phase front at which AFE, FE, and intermediate phases coexist, revealing a transient and spatially inhomogeneous transformation pathway. This direct, real-time observation of mixed and electric-field-dependent FE and AFE order reveals how competing structural instabilities govern the AFE–to-FE transition, providing an atomistic blueprint for controlling such phase transformations under device-relevant conditions.

\section{Atomic-Scale Phase Transition Pathway}

The investigated thin-film capacitor devices are comprised of 100-nm-thick \hkl(240)\textsubscript{O}-oriented PZO thin films (subscript O denotes orthorhombic indices) sandwiched between \ce{SrRuO3} (SRO) top and bottom electrodes, all grown on a conductive Nb-doped \ce{SrTiO3} (Nb:STO) substrate. Samples for \textit{operando} STEM biasing are prepared using a focused-ion beam and dedicated micro-electromechanical systems (MEMS) chips (Figure~\ref{fig:overview_pzo_struct}a, details in Supplementary information, Section 1) \cite{Pan2024-fr}. Notches are cut into the top platinum contact and bottom substrate to enable application of the electric field along the PZO film out-of-plane direction, consistent with the operation of bulk thin-film capacitors.

In the absence of electrical bias, \nicefrac{1}{4} 110\,\textsubscript{PC} (subscript PC denotes pseudocubic indexing)  superlattice reflections are present in the nano-beam electron diffraction (NBED) patterns across the PZO films (Figure~\ref{fig:overview_pzo_struct}b). These reflections arise from the anti-parallel lead displacements (of the form: $\uparrow\uparrow\downarrow\downarrow$) characteristic of the AFE ground state \cite{Sawaguchi1951-xy}. In-plane rotational variants, induced by the isotropic lattice of the Nb:STO substrate, are visualized in the color-mixed dark-field images formed using the \nicefrac{1}{4} 110-type reflections (Figure~\ref{fig:overview_pzo_struct}b-c), where yellow and red regions correspond to domains with in-plane mirrored antipolar axes (indicated with colored arrows). In contrast, dark regions are consistent with domains in which the antipolar axis is oriented parallel to the view direction, resulting in the absence of superlattice reflections.

At the atomic scale, simultaneously-acquired annular dark field (ADF) and differentiated differential phase contrast (dDPC) images (Figure~\ref{fig:overview_pzo_struct}d-e, details in Supplementary Information, Section 1) acquired from a red-region further confirm the AFE ground state. The overlaid lead–oxygen polar displacements (Figure~\ref{fig:overview_pzo_struct}d, details in Supplementary Information, Section 1) reflect the characteristic $\uparrow\uparrow\downarrow\downarrow$ ordering along \hkl<110>\textsubscript{PC}, while the oxygen-zirconium-oxygen angles show antiferrodistortive rotation of the oxygen octahedra (Figure~\ref{fig:overview_pzo_struct}e). Both structural features share the same periodicity and are consistent with the atomic model of orthorhombic AFE (AFE\textsubscript{O}) PZO (inset, Figure~\ref{fig:overview_pzo_struct}d-e \cite{Sawaguchi1951-xy,Teslic1998-ba}).

The field-driven structural evolution is subsequently tracked through simultaneous atomic-resolution ADF and dDPC by maintaining the same field-of-view throughout a complete bias cycle (\textit{i.e.}, the bias path is 0 V $\rightarrow$ +8 V $\rightarrow$ -7 V $\rightarrow$ 0 V, where 1 V is equivalent to 100 kV/cm given the 100-nm-thick film). Four distinct phases emerge sequentially with increasing applied field strength, and their corresponding atomic motifs are extracted by averaging regions with similar structural features (Figure~\ref{fig:atom_motifs}, details in Supplementary Information, Section 1). The first stage in the sequence corresponds to the orthorhombic AFE ground state (AFE\textsubscript{O}, Figure~\ref{fig:atom_motifs}a), characterized by $\uparrow\uparrow\downarrow\downarrow$ antipolar displacements coupled with antiferrodistortive oxygen octahedral rotation. When a sufficiently large external field field is applied, both structural features are suppressed synchronously, ultimately leading to the rhombohedral FE phase with \hkl[111]-aligned dipoles (FE\textsubscript{R} in Figure~\ref{fig:atom_motifs}c) and consistent with bulk X-ray studies \cite{Liu2020-at, Blue1996-po, Park1997-sb}. Beyond this critical field, the FE\textsubscript{R} phase behaves as a conventional ferroelectric, where increasing field strength further aligns dipoles to the applied field direction (\hkl[001]\textsubscript{PC}) and induces a tetragonal phase, similar to \ce{BaTiO3} (FE\textsubscript{T}, Figure~\ref{fig:atom_motifs}d) \cite{Strukov2011-ee, Zhao2016-ni}.

Between the AFE\textsubscript{O} and FE\textsubscript{R} phases, a bridging structure emerges with monoclinic-type superlattice reflections in its discrete Fourier transform. These reflections are reminiscent of those observed at the AFE/FE boundary in \ce{PbZr$_{1-x}$Ti$_x$O$_3$} (PZT), and have been attributed to antiparallel lead displacements (M\textsubscript{Pb}-type, $\uparrow\downarrow\uparrow\downarrow$) and octahedral rotations (M\textsubscript{O}-type, $a^0a^0c^+$) of an intermediate phase \cite{Fu2022-vn}. In the present films, the overlaid octahedral rotations on the dDPC image show a checkerboard modulation that coincides with the M\textsubscript{O}-type rotations, and residual M\textsubscript{Pb}-like ordering persists as a checkerboard pattern of canted AFE-like displacements superimposed on top of the mostly aligned FE-like dipoles (mean-subtracted inset, Figure~\ref{fig:atom_motifs}b).

Similar intermediate states also appear in PZO films near the critical thickness for emergence of the FE phase ($\sim$25 nm) \cite{Jiang2024-db}. Despite differing origins—thickness \cite{Jiang2024-db}, titanium alloying \cite{Fu2022-vn}, or an electric electric field as applied here—these phases consistently appear under a common condition: competition between AFE and FE order. This reveals a shared energetic origin, in which structural constraints, composition, or applied field modulate the free-energy landscape of structural instabilities. As a result, the intermediate phase exhibits a superposition of both AFE and FE structural characteristics, providing a low-energy pathway that facilitates the phase transformation \cite{Fu2022-vn, Si2025-cn}, while the same checkerboard periodicity reinforces the strong coupling between dipolar cation displacements and antiferrodistortive octahedral rotations during the process. 

% Part of the DFT results can come in here.

Together, the four identified phases (AFE\textsubscript{O} $\rightarrow$ FE\textsubscript{M} $\rightarrow$ FE\textsubscript{R} $\rightarrow$ FE\textsubscript{T}) establish an atomic-scale pathway for the AFE-to-FE transition in PZO under device relevant conditions. In particular, the coexistence of AFE-like modulations within the polar FE\textsubscript{M} matrix indicates a stepwise transformation mechanism, governed in part by competing structural instabilities under changing external bias \cite{Yu2024-sg, Jiang2023-gp, Singh1995-yj, Roy-Chaudhuri2011-ew}.

\section{Local Inhomogeneities in Phase Stability}

\begin{figure}[!h]
    \centering
    \includegraphics[width=6.5in]{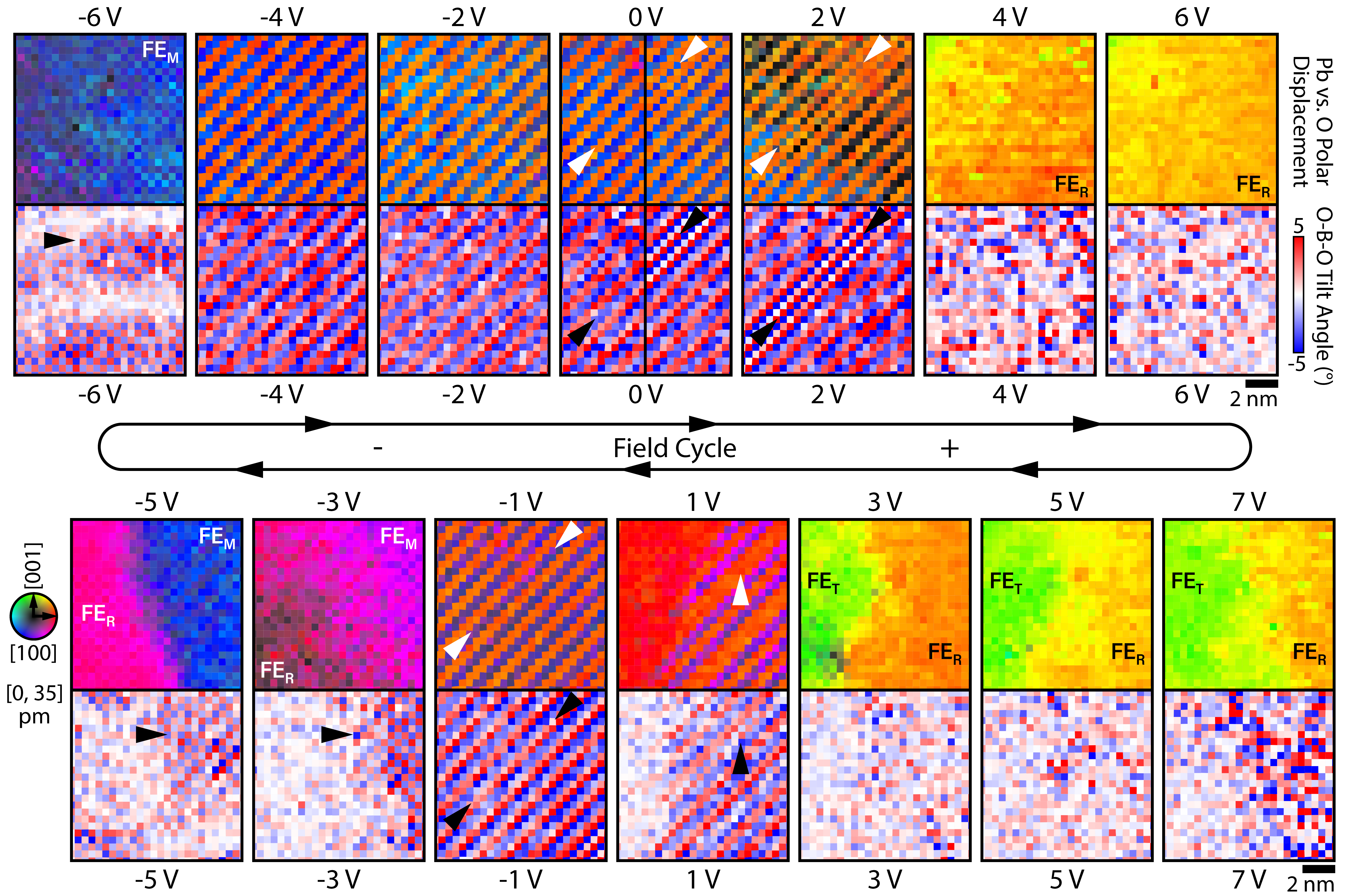}
    \caption{Lead vs. oxygen polar displacements and octahedral rotation or tilt angle extracted for each lead-centered unit cell over the same field of view as a function of biasing voltage. Arrows indicate translation boundaries at low applied biases.  At -3, -5, and -6 V, arrows also indicate the presence of a checkerboard octahedral rotation pattern associated with the FE$_{M}$ phase.}
    \label{fig:atom_surface_cyling}
\end{figure}

Despite the shared stages in the AFE-to-FE transition, local perturbations \cite{Liu2023-wg}, such as dislocations \cite{Yu2024-sg, Liu2025-jj}, point defects \cite{Wei2021-mp}, strain gradients \cite{Jiang2024-db, Roy-Chaudhuri2011-ew}, or relaxation of substrate clamping \cite{Acharya2023-pv, Pan2024-fr}, can alter the local energetics and therefore the stability of the phases. As a result, long-range lattice order can be disrupted, giving rise to heterogeneity and non-uniform transformations. This inhomogeneity becomes evident when looking beyond the single unit-cell motifs in the atomic-resolution images. For example, imaging over the same $\sim$ 10 $\times$ 10 nm$^2$ field-of-view throughout the full \textit{operando} biasing cycle (Figure~\ref{fig:atom_surface_cyling}) enables direct atom-column-by-atom-column tracking of field-induced structural changes. At the ground state (0 V, right half of panel, Figure~\ref{fig:atom_surface_cyling}) the film structure is predominantly consistent with the AFE\textsubscript{O} phase with several translation boundaries (white arrows) disrupting the long-range AFE-dipole and antiferrodistortive-octahedral order \cite{Yu2024-sg, Jiang2024-db, Wei2014-ir}. These translation boundaries, also visible as diagonal lines in NBED dark-field images (arrows in Figure~\ref{fig:overview_pzo_struct}c), have been previously associated with a strain-mediated ferrielectric (FiE) phase \cite{Yu2024-sg}. 

As the applied electric field increases, the dipoles on one polar sub-lattice decrease in magnitude and gradually rotate to align with the opposing sub-lattice, all while preserving the original translation boundaries. Beyond a critical threshold, between +2 V and +4 V, this configuration transforms to a unidirectional polar matrix consisting of FE\textsubscript{M} and FE\textsubscript{R} phases, completing the AFE-to-FE transformation. Increasing the field further drives part of the region into the FE\textsubscript{T} phase (green region at +7 V). Upon field removal, the structural changes reverse, retracing the transition pathway. During the reverse cycle (0 V $\rightarrow$ -7 V $\rightarrow$ 0 V), polar displacements mirror those aforementioned, reorienting with the applied field towards the bottom electrode.

While the transition generally aligns with previously identified atomic-scale motifs (\textit{c.f.} Figure~\ref{fig:atom_motifs}), pronounced inhomogeneity is evident. The full path of the transformation exhibits asymmetric hysteresis behavior \cite{Randall2021-xo}. For example, the AFE state in the as-prepared sample (0 V) is not fully restored until -1 V (0 V $\rightarrow$ +7 V $\rightarrow$ -1 V). Furthermore, upon completing the cycle (-1 V $\rightarrow$ -6 V $\rightarrow$ 0 V), the original translation boundaries observed prior to biasing are absent, and the AFE order vector is fully out-of-phase from that of the -1 V AFE state (opposite polar displacement for the same plane indicated by the arrow, Figure~\ref{fig:atom_surface_cyling}). These observations offer proof that coercivity is not solely governed by domain-wall motion or macroscopic pinning effects, but can also manifest through sub-lattice polarization shifts and local structural distortions at the atomic scale \cite{Fengler2017-hb}. Residual FE ordering, whether in the form of translation boundaries or uncompensated AFE dipoles, thus contributes to the larger-than-bulk remanent polarization frequently observed in thin films \cite{Si2025-cn}.

Spatial inhomogeneity of the phase transformation in response to the field is evident as well: the left half of the probed region exhibits higher susceptibility, manifested by a lower threshold for the AFE-to-FE phase transition (\textit{c.f.} +1 V at bottom half, Figure~\ref{fig:atom_surface_cyling}) and stronger dipole alignment with the applied field in the FE state (\textit{c.f.}, +7 V). This trend persists throughout the entire biasing cycle and reflects defect-modulated variations in the energy landscape \cite{Yu2024-sg, Li2025-yw, Wei2021-mp}. In contrast, the emergence of FE\textsubscript{M} and FE\textsubscript{R} phases, characterized by checkerboard octahedral rotations, fluctuates spatially without any apparent trends, especially at -6 V. This stochastic behavior can arise from the close energetic proximity between the two phases, where small perturbations can induce interconversion (\textit{e.g.}, due to local strain, thermal fluctuations, or electric field inhomogeneity). Ultimately, the inhomogeneity imposed by the thin-film geometry, combined with the intrinsic role of the intermediate FE\textsubscript{M} phase, accounts for the slanted hysteresis loop that deviates from ideal AFE behavior \cite{Si2025-cn}.

\section{Depth-Dependent Film Response}

Beyond local defect-induced modulation of the film response, longer-range depth-dependent effects arising from relaxation of substrate clamping/strain or free surfaces can also modulate the energy landscape \cite{Acharya2023-pv}. For example, a FE ground state has been reported in PZO films thinner than 15-25 nm \cite{Yu2024-sg, Roy-Chaudhuri2011-ew, Jiang2024-db} as compared to the 100-nm-thick PZO film investigated here. Moreover, those observations in thin PZO films, however, are limited to comparisons across films of varying thicknesses, leaving it unclear how these long-range effects may interact within a film of fixed thickness during external bias.

\begin{figure}[!h]
    \centering
    \includegraphics[width=3in]{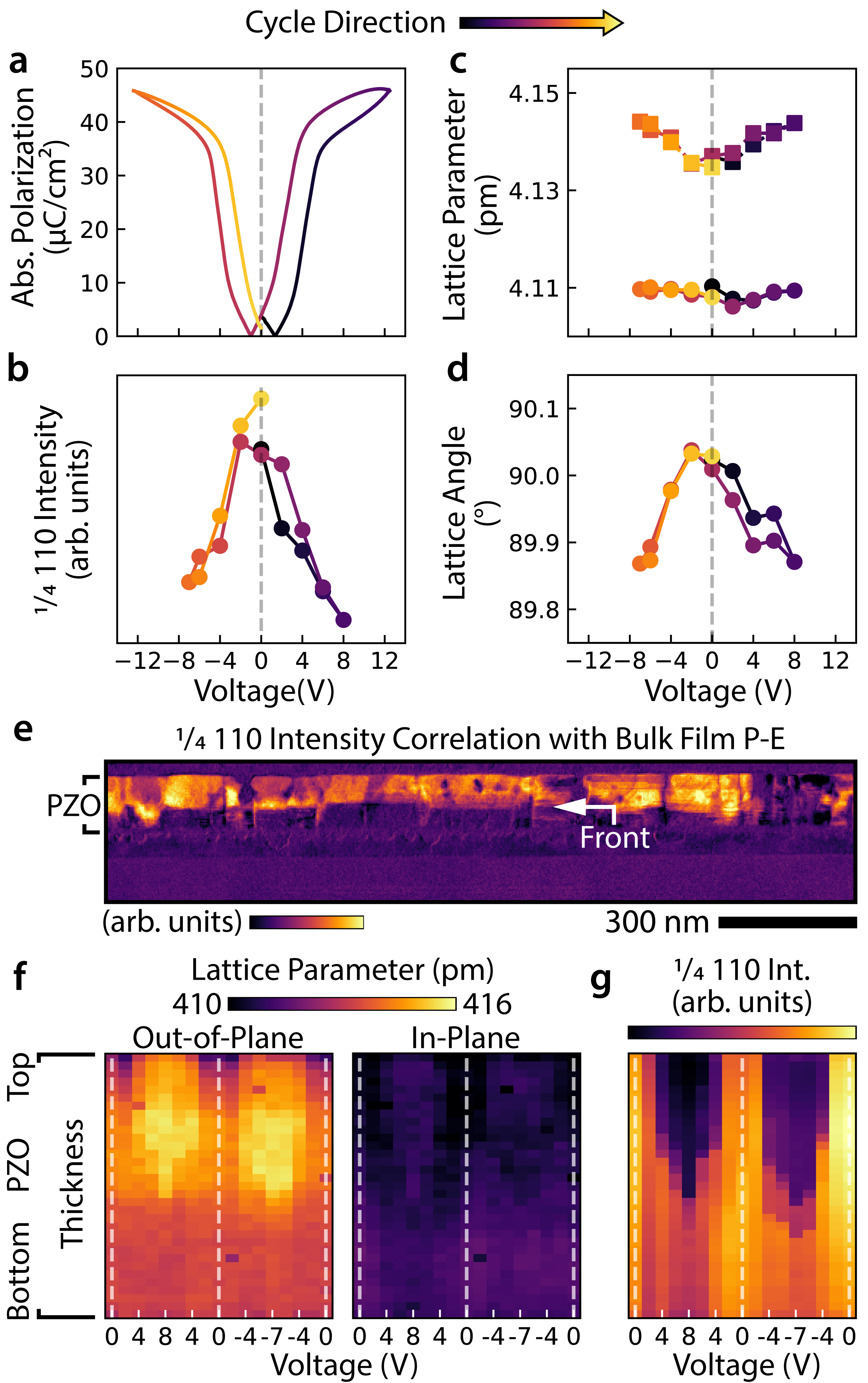}
    \caption{(a) Magnitude of polarization, (b) lattice angle, (c) intensity of the \nicefrac{1}{4} 110 superlattice reflection, and (d) lattice parameters as a function of applied voltage, where (a) is measured macroscopically from the device, and (b–d) are measured using NBED patterns. (e) Correlation between voltage-dependent superlattice intensity and macroscopic polarization magnitude, reflecting the similarity with macroscopic behavior at each scan position. Thickness-resolved change in (f) lattice parameters and (g) Intensity of \nicefrac{1}{4} 110 reflections as a function of applied voltage.}
    \label{fig:overview_pzo_struct_cycling}
\end{figure}

To probe this depth-dependence at the film scale, the field-dependent structural evolution of the \textit{operando} samples is first compared to the bulk capacitor polarization response (Figure~\ref{fig:overview_pzo_struct_cycling}a) using NBED and cepstral analysis \cite{Padgett2020-nl}. With increasing applied field, the collapse of AFE order is marked by a progressive decrease in the \nicefrac{1}{4} 110 superlattice reflection intensity (Figure~\ref{fig:overview_pzo_struct_cycling}b). Concurrently, the average out-of-plane lattice parameter increases, while the in-plane lattice parameter remains constrained by the substrate (Figure~\ref{fig:overview_pzo_struct_cycling}c) \cite{Pan2024-fr}. The lattice angle deviates from its near-orthogonal ground-state configuration as well ($\sim$ 90°), aligning with the AFE\textsubscript{O}-to-FE\textsubscript{R} transition (Figure~\ref{fig:overview_pzo_struct_cycling}d). The expansion of the lattice parameter along the field direction and deviation from orthogonality correlate with increasing polarization magnitude ($\lvert$P$\rvert$) at higher electric field in bulk capacitors (Figure~\ref{fig:overview_pzo_struct_cycling}a), confirming the field-driven phase transition. 

Leveraging the spatially resolved NBED patterns from 4D-STEM, the spatial variation of the transition is evaluated across the film by correlating the voltage-dependent \nicefrac{1}{4} 110 superlattice intensity at each scan position with the bulk $\lvert$P$\rvert$-E measurements. This metric highlights regions of the film that either respond consistently with, or deviate from, the macroscopic transition (bright and dark regions respectively, Figure~\ref{fig:overview_pzo_struct_cycling}e). Aside from minor domain-to-domain variations, a systematic depth dependence is observed \cite{Jiang2024-db}. Regions near the top electrode exhibit structural responses consistent with bulk behavior, including a hysteretic decrease in superlattice intensity under increasing field. In contrast, the structural changes near the bottom electrode show almost no correlation with the bulk-device response. 

To further highlight the structural variations, the lattice parameters and superlattice-reflection intensities along the film-growth direction are averaged and mapped as a function of the applied field (Figure~\ref{fig:overview_pzo_struct_cycling}f-g), revealing two defining features. The superlattice-reflection intensity at 0 V is notably reduced in regions near the bottom electrode, signaling weaker AFE order relative to regions further away in the ground state. Moreover, this suppressed AFE state remains largely pinned under bias, as evidenced by smaller variations in the lattice parameter and superlattice reflection intensity.

The behavior near the substrate reveals an effective ``dead layer'' below a critical thickness, contrasting with the more active behavior observed farther from the interface. Atomic-resolution imaging from different regions of the film (Supplementary Information, Figure S1) provides further evidence of this depth-dependent heterogeneity. The dead layer persists near the substrate independent of the biasing direction and the symmetric electrode configuration, ruling out an electric field gradient or surface depolarization as its primary origin \cite{Mani2015-iv}. Instead, spatially varying dislocation density and partial relaxation of substrate clamping \cite{Yu2024-sg, Li2025-yw, Wei2021-mp, Jiang2024-db, Roy-Chaudhuri2011-ew, Acharya2023-pv} can generate an internal field that disrupts structural uniformity, modulates the film response across the thickness, and reshapes the energy landscape governing phase stability. By counteracting the applied electric field, this effective internal field contributes to a depth-dependent non-uniform phase transition, while local defects further amplify this heterogeneity. Thus, understanding the interplay between the heterogeneous internal field and the external applied field are essential to optimizing the device’s functional properties.

\section{Probing Dynamic Phase Equilibria}

In addition to the near-instantaneous phase transformation, pronounced heterogeneity poses further challenges for experimentally capturing the transformation dynamics \cite{Cai2003-mp, Blue1996-po, Kajewski2020-gl, Jiang2024-db, Wei2021-mp}. The depth-dependent film response observed here, however, suggests the existence of a transition front (horizontal streaks indicated in Figure~\ref{fig:overview_pzo_struct_cycling}e and Supplementary Information, Figure S2) that offers a unique opportunity to observe phase dynamics under operating conditions.

Atomic-resolution ADF and dDPC imaging at -6 V (frames in Figure~\ref{fig:flicker}a-d and full series in Supplementary Information, Video S1) resolves coexisting AFE and FE phases at this front, separated by a defined boundary (dashed line) and accompanied by translation boundaries and AFE rotational variants. This coexistence arises from the competition between the spatially graded internal field (strengthening toward the bottom electrode due to relaxation of clamping and dislocation gradients) and the uniform external applied field. At the front, these competing effects flatten the energy landscape among different phases, resulting in their metastable coexistence.

Below the transition front, the progressively stronger internal field stabilizes the low-energy AFE\textsubscript{O} ground state. This region also hosts translation boundaries that act as intermediate states (Figure~\ref{fig:flicker}d and Supplementary Information, Video S1), supporting non-zero polarization and nucleating the FE phase prior to the full transition\cite{Xu2024-aw}. Above the front, FE phases dominate, consistent with their higher ground-state energy relative to AFE\textsubscript{O}. Additionally, a checkerboard pattern of octahedral rotations emerges, gradually diminishing in magnitude with increasing distance (Figure~\ref{fig:flicker}c and Supplementary Information, Figure S3). This trend represents a continuous progression from the FE\textsubscript{M} to the FE\textsubscript{R} phase (\textit{cf.} Figure~\ref{fig:atom_motifs}) as internal fields weaken away from the transition front. In other words, FE\textsubscript{M} exhibits a lower relative energy when compared to FE\textsubscript{R}. The reduced energy of FE\textsubscript{M} is consistent with its antiferrodistortive rotations and antipolar canted cation displacements along \hkl[1-10] \cite{Jiang2024-db}, both of which represent residual AFE-like order and suggest its closer energetic proximity with the AFE\textsubscript{O} phase.

Over successive frames, the metastable transition front and translation boundaries move dynamically (Figure~\ref{fig:flicker}a and Supplementary Information, Video S1), further indicating the near-degenerate energies among different phases in this region \cite{Liu2023-uh, Yu2024-sg, Jiang2023-gp, Singh1995-yj, Roy-Chaudhuri2011-ew}. Within this flattened energy landscape, beam-induced charging and heating effects can transiently disturb the phase boundary, initiating localized phase transitions. In turn, these transitions further propagate polarization and electric-field variations, dynamically altering boundary conditions for adjacent regions. This results in oscillations between phases and prevents the system from reaching a local static equilibrium. Although imaged with an electron beam, the stochastic and reversible evolution, without consistent directional progression, persists and contrasts with the directed phase transitions induced by prolonged electron-beam irradiation in prior studies \cite{Wei2020-gu, Wei2021-mp, Calderon2023-cv, Calderon2025-yt}. Instead, the observed dynamics are governed by the intrinsic properties of the film and the applied electric field, reflecting the \textit{operando} film response to local perturbations (as further confirmed by a lower-dose series in Supplementary Information, Figure S4 and Video-S1).

The spatial profile of the internal field can be approximated by monitoring the average phase transition front position as a function of applied electric fields using Fourier filtering (Figure~\ref{fig:flicker}e-g, details in Supplementary Information, Section 4). Hysteresis behavior is evident during cycling, as the front shifts incrementally in the direction of the increasing external field. Under a 4 V potential step (4 V to 8 V, or equivalently a field change of 400 kV/cm), the transition front advances by only 30–40 nm. Such a limited shift under a large change in applied field suggests that the effective internal field is large and spatial limiting within the field-of-view, \textit{e.g.}, step-like profile from the bottom to the top electrode. Beyond this critical region, phase transitions can proceed to completion with minimal pinning effects.

The standard deviation of the transition front position (metastability width) at fixed applied fields provides a gauge for the phase stability. The front width peaks at intermediate biases (6 V), where the energy landscapes of different phases converge, maximizing the system's sensitivity to perturbations and thus the extent of spatial fluctuation. Furthermore, the width increases under higher electron beam doses at a given field (Supplementary Information, Figure S4), consistent with enhanced perturbation-driven destabilization of the phase boundary. In contrast, at lower (4 V) and higher (8 V) biases, the region is characterized by AFE- or FE-dominated states, respectively, suppressing the magnitude of fluctuations as the energy balance favors a single-phase configuration.

Overall, the observed phase coexistence is analogous to phase-field simulations of heterogeneously doped bulk PZO, where a doping-induced rise and fall in local potential creates polar boundaries that act as nucleation sites for AFE-to-FE transitions at reduced critical fields, stabilizing phase coexistence \cite{Xu2024-aw}. Although local polar phases (translation boundaries) are also observed in the thin films here, the relaxation of clamping and dislocation gradients near the film-substrate interface dominate the energy modulation, establishing a transition front governed by the interplay of internal fields and the applied external field. Furthermore, perturbations near the front, such as electron-beam-induced fluctuations as discussed earlier, yield dynamic phase coexistence, distinct from the static interfaces in the simulations. Despite differing origins (doping vs. clamping/dislocations) and temporal behavior (steady vs. dynamic), both mechanisms highlight the critical role of local heterogeneities in reshaping energy landscapes and governing phase transition pathways.

\section{Conclusions}

Enabled by \textit{operando} STEM under device-relevant conditions, this study provides a direct, atomic-scale determination of the multi-step AFE-to-FE phase transition in PZO thin films. The transition follows four distinct structural motifs: the AFE ground state (AFE\textsubscript{O}) first evolves into a transient monoclinic ferroelectric phase (FE\textsubscript{M}), stabilizes into a rhombohedral FE phase (FE\textsubscript{R}), and ultimately adopts conventional ferroelectric behavior with tetragonal distortion (FE\textsubscript{T}) at higher fields. This evolution is further modulated by pronounced spatial and temporal heterogeneity, as revealed through NBED and atomic-resolution imaging. A depth-dependent clamping relaxation and dislocation distribution generate an internal field that competes with the applied field. This leads to a ``dead layer'' near the bottom electrode, with more dynamic and bulk-like switching occurring near the top electrode. Local defects amplify this complexity, introducing region-to-region variability in phase stability. Finally, AFE and transient FE phases coexist at the phase front and are stabilized by competing depth-dependent heterogeneities, enabling direct atomic-scale observation of phase transformation dynamics under \textit{operando} conditions. These findings provide fundamental insights into field-driven phase transitions in PZO thin films and underscore the critical role of heterogeneities and defect engineering in shaping the functional properties of antiferroelectric devices.

\newpage
\begin{figure}[!h]
    \centering
    \includegraphics[width=3in]{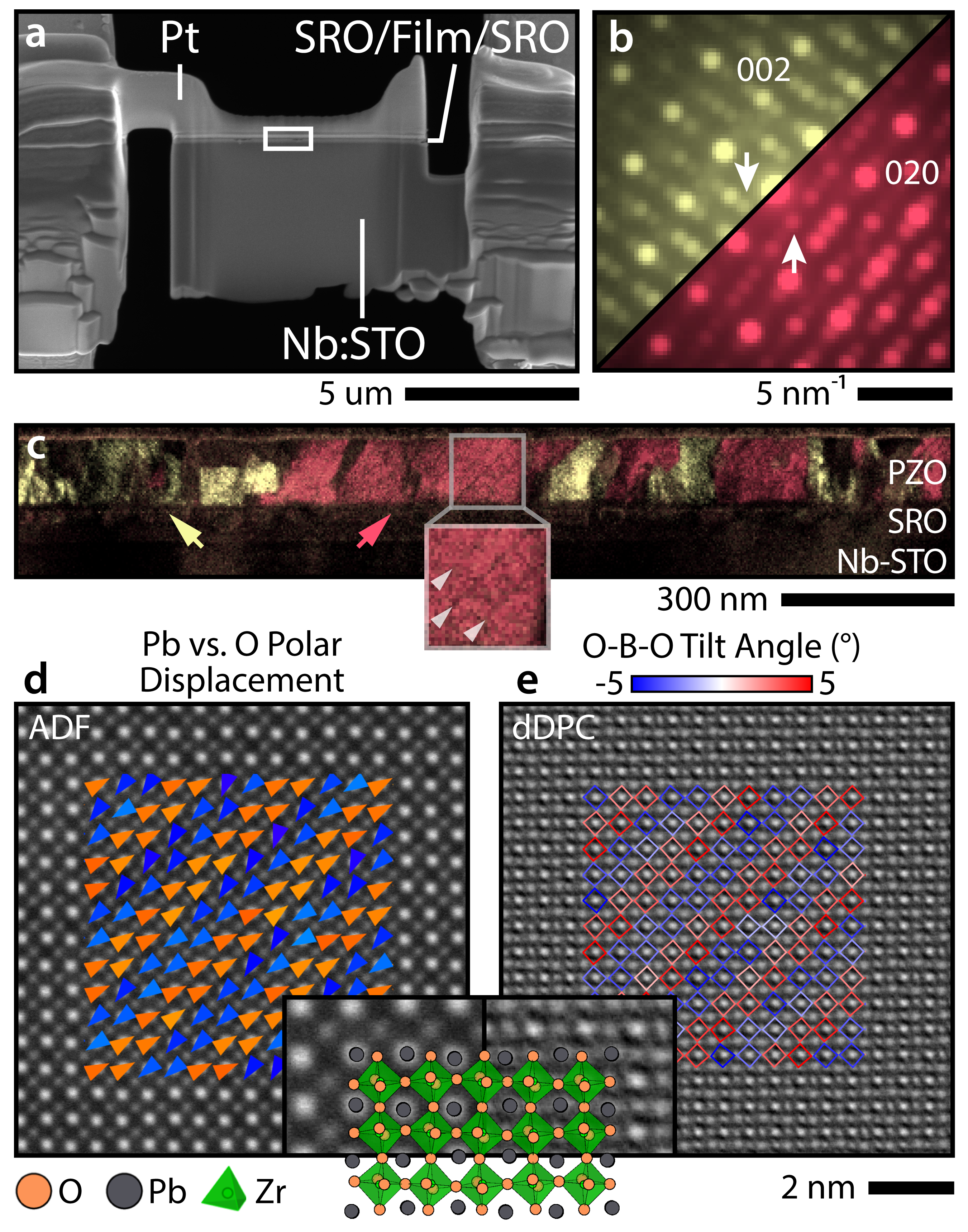}
    \caption{(a) Overview of the STEM \textit{operando} biasing setup with labeled device layers. (b) Averaged diffraction pattern from two AFE domains rotated in-plane, with \nicefrac{1}{4} 110 superlattice reflections indicated by arrows. (c) Real-space distribution of the two AFE domains within the boxed region in (a), visualized with color-mixed dark-field images formed from the superlattice reflections indicated in (b). (d) Atomic-resolution ADF image of the AFE PZO overlaid with polar displacements. (e) dDPC image from the same region overlaid with octahedral rotation or tilt angles. The inset shows the images with a crystal model of AFE PZO.}
    \label{fig:overview_pzo_struct}
\end{figure}

\newpage
\begin{figure}[!h]
    \centering
    \includegraphics[width=3in]{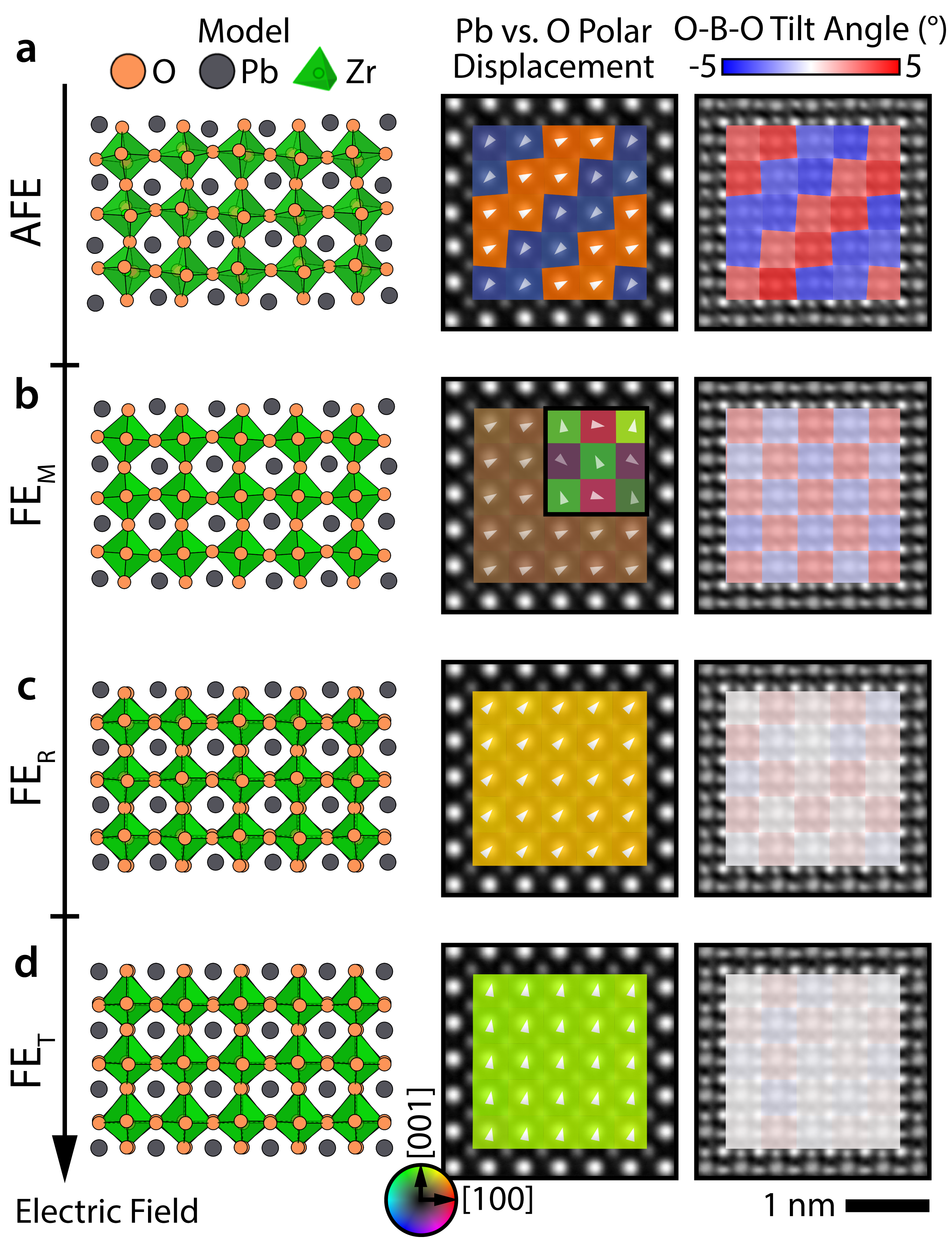}
    \caption{Crystal models and motifs of (a) AFE, (b) FE$_{M}$, (c) FE$_{R}$, and (d) FE$_{T}$ phases obtained from experiment by averaging ADF and dDPC images from atomic resolution images during cycling. The lead vs. oxygen polar displacements and oxygen octahedral rotation or tilt angles are overlaid. The mean-subtracted displacements for the FE$_{M}$ motif is shown in the inset in (b), with a saturation range of [0, 2] pm.}
    \label{fig:atom_motifs}
\end{figure}

\newpage

\newpage

\newpage
\begin{figure}[!h]
    \centering
    \includegraphics[width=6.5in]{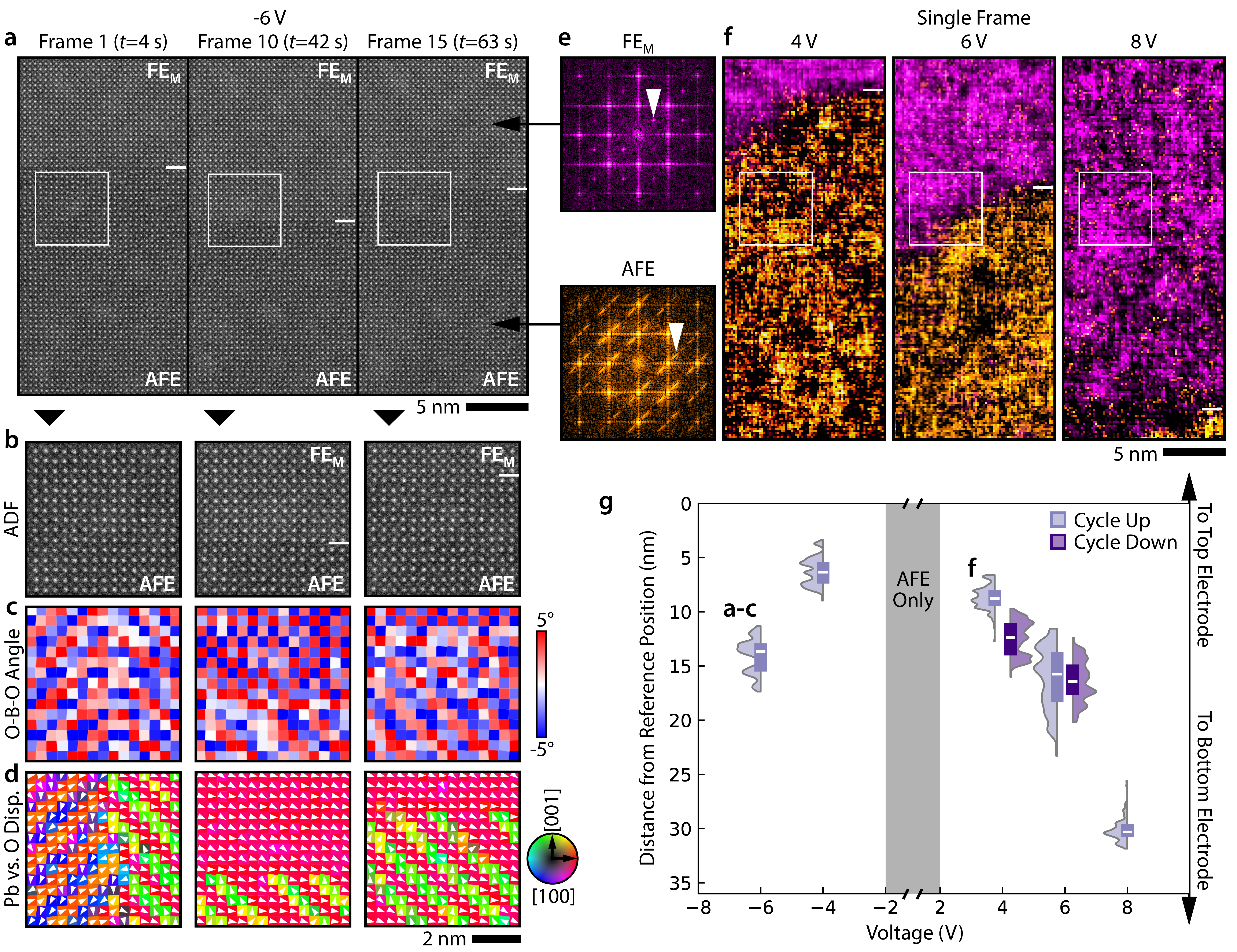}
    \caption{(a) Sequence of images collected from the same region under -6 V showing the metastable phase transitions between AFE and FE phases as indicated by dashed lines. (b) Magnified view of the boxed region in (a), along with (c) octahedral rotation patterns and (d) lead polar displacement patterns. (e) Discrete Fourier transform from the FE and AFE regions, with the \nicefrac{1}{4} 110 superlattice reflection observable only for the AFE phase, as indicated by the arrows. (f) The intensity of the AFE \nicefrac{1}{4} image frequency is extracted by Fourier filtering of ADF images acquired from the same registered region under 4, 6, and 8 V applied bias. False-color mapping highlights the different phases in each region, revealing the directional movement of the transition front. (g) Violin plot showing the distribution of the transition front within the sequential images at each applied voltage, with the width of the violin representing the density of the distribution. The 25th and 75th percentiles are indicated by the colored bar, while the median or 50th percentile is marked by the white line.}
    \label{fig:flicker}
\end{figure}

\newpage
\bibliography{paperpile.bib}

\newpage
\section*{Acknowledgments}
This research was sponsored by the Army Research Laboratory and was accomplished under Cooperative Agreement Number W911NF-24-2-0100. The views and conclusions contained in this document and those of the authors should not be interpreted as representing the official policies, either expressed or implied, of the Army Research Laboratory or the U.S. Government. The U.S. Government is authorized to reproduce and distribute reprints for Government purposes, notwithstanding any copyright notation herein. L.W.M. and J.M.L. acknowledges additional support from the Air Force Office of Scientific Research under award number FA9550-24-1-0266. Any opinions, findings, and conclusions or recommendations expressed in this material are those of the author(s) and do not necessarily reflect the views of the United States Air Force. This work made use of the MIT.nano Characterization Facilities. 

\section*{Author Information} 
\subsection*{Contributions}
M.Z. and M.X. proposed the experiment design and contributed equally to this work. M.Z., M.X., and C.G. performed sample preparation, electron microscopy, and data analysis. H.P. synthesized the thin films. L.A. provided theory contributions and structure models. J.M.L., L.W.M., and G.H supervised the research. All authors co-wrote and edited the manuscript.

\section*{Ethics declarations}
\subsection*{Competing Interests}
The authors declare no competing interests.

\section*{Data Availability} 
Data supporting the study are available from the corresponding author on reasonable request. 

\end{document}